\documentstyle[aps,prl,multicol,epsf]{revtex}
\begin{document} 
\draft

\title{Topological Scenario for Stripe Formation in Manganese Oxides}

\author{Takashi Hotta$^{1,2}$, Yasutami Takada$^{2}$,
Hiroyasu Koizumi$^{3}$, and Elbio Dagotto$^{1}$}

\address{$^1$ National High Magnetic Field Laboratory,
Florida State University, Tallahassee, FL 32306}

\address{$^2$ Institute for Solid State Physics,
University of Tokyo, 7-22-1 Roppongi, Minato-ku, 
Tokyo 106-8666, Japan}

\address{$^3$ Faculty of Science, Himeji Institute of Technology,
Kamigori, Ako-gun, Hyogo 678-1297, Japan}

\date{August 13, 1999}

\maketitle

\begin{abstract}
The spin-charge-orbital complex structures of manganites are 
studied using topological concepts.
The key quantity is the ``winding number'' $w$ associated with the 
Berry-phase connection of an $e_g$ electron parallel-transported 
through Jahn-Teller centers, along zigzag one-dimensional paths 
in an antiferromagnetic environment of $t_{2g}$ spins. From 
these concepts, it is shown that the ``bi-stripe'' and ``Wigner-crystal'' 
states observed experimentally have different $w$'s.
Predictions for the spin structure of the charge-ordered states for 
heavily doped manganites are discussed.
\end{abstract}


\begin{multicols}{2}
\narrowtext

The curious static patterns in the spin, charge, and orbital 
densities observed in manganites are currently attracting much 
attention \cite{Cheong}.
In La$_{1-x}$Ca$_{x}$MnO$_{3}$, the CE-type antiferromagnetism 
(AFM) has been observed at $x$=$1/2$ since the 1950's \cite{Wollan},
but recently a similar structure has been proposed at $x$=$2/3$ based 
on neutron diffraction experiments \cite{Radaelli}. 
In these AFM structures, the $t_{2g}$ spins align in parallel 
along zigzag-shaped one-dimensional (1D) paths in the $a$-axis 
direction, while they are antiparallel across these paths, which are
stacked in the $b$-axis direction. 
This spin arrangement here is called the ``stripe-AFM''(S-AFM) structure.

Interestingly enough, the charge and orbital ordering (COO)
observed experimentally is always concomitant 
with this S-AFM phase. 
At $x$=$1/2$, the COO has been confirmed by the synchrotron X-ray 
diffraction experiment \cite{Murakami}. 
At $x$=$2/3$, however, two different COO patterns have been 
reported; the ``bi-stripe'' (BS) structure \cite{Mori}, 
in which the main building block of the COO pattern at $x$=$1/2$ 
persists even at $x$=$2/3$, and the ``Wigner-crystal'' (WC) 
structure \cite{Radaelli}, in which the COO occurs 
with the distance between charges maximized. 
The appearance of the two different structures indicates that 
(1) the corresponding energies are very close to each other, namely, 
the ground state has (quasi-)degeneracy and that (2) the conversion 
between them is prohibited by a large energy barrier.
Under these circumstances, it is of limited relevance 
the determination of which of the two states is 
better energetically based on some model Hamiltonian.

In such a subtle situation, 
we focus on the origin of the near BS-WC degeneracy,
rather than calculating which one is the true ground state.
We follow the strategy of labeling these 
(quasi-)degenerate ground states in terms of a physically-motivated 
quantity which does not necessarily manifest itself in the Hamiltonian $H$. 
Specifically we focus our attention on the important role of the 
1D conducting zigzag paths in the $a$-$b$ basal plane, and 
considered parallel-transport of an $e_g$ electron along these paths
through the Jahn-Teller (JT) centers composed of MnO$_6$ octahedra. 
The transport invokes the Berry-phase connection and we can 
introduce the ``winding number'' $w$ as a direct consequence of 
topological invariance which should be conserved irrespective 
of the details of $H$.

In this Letter, we propose that such a topological invariance is a key 
concept to understand the complex states of manganites 
since we observe that $w$ is always a good index 
to label the (quasi-)degenerate S-AFM states for $x$$\ge$1/2.
In fact, it is found that if the energies for various paths considered here 
are plotted, its distribution contains a multifold-band structure 
indexed by $w$.
The observed WC and BS structures belong to 
two topologically different classes, characterized by the 1D paths 
with $w$=1 and $w$=$x$/(1$-$$x$), respectively.
The conversion between the WC and BS states is hindered by a 
large energy barrier due to this topological difference.

Consider $e_g$ electrons coupled to both localized $t_{2g}$ spins,
with the Hund's rule coupling $J_{\rm H}$, and JT distortions of 
the MnO$_6$ octahedra.
Since it is the largest characteristic energy among those considered here,
$J_{\rm H}$ is taken to be infinite for simplicity. 
This implies that the spin of each $e_g$ electron at a Mn site 
aligns completely in parallel with the direction of the $t_{2g}$ spins 
at the same site. 
Thus, the spin degrees of freedom are effectively lost for the 
$e_g$ electrons, and the spin index will be dropped hereafter. 
Since experimentally it is known that the $t_{2g}$ spins are 
antiparallel along the $c$-axis, we can assume that the $e_g$ electrons 
can move only in the $a$-$b$ plane.

The above situation is well described in terms of $H$ as 
\begin{eqnarray}
 \label{hamiltonian}
 H &=& - \! \sum_{{\bf ia}\gamma\gamma'}
 t^{\bf a}_{\gamma \gamma'} d_{\gamma{\bf i} }^{\dag} d_{ \gamma'{\bf i+a}}
 \!+\! J' \sum_{\langle {\bf i,j} \rangle}
 {\bf S}_{\bf i} \cdot {\bf S}_{\bf j} \nonumber \\
 &+& \!E_{\rm JT} \! \sum_{\bf i} \Bigl[
 2(q_{2{\bf i}} \tau_{x{\bf i}} + q_{3{\bf i}} \tau_{z{\bf i}})
 +(q_{2{\bf i}}^2+q_{3{\bf i}}^2) \Bigr],
\end{eqnarray}
with $\tau_{x{\bf i}}$=$d_{a{\bf i}}^{\dag}d_{b{\bf i}}$+
$d_{b{\bf i}}^{\dag} d_{a{\bf i}}$ and 
$\tau_{z{\bf i}}$=$d_{a{\bf i}}^{\dag}d_{a{\bf i}}$$-$
$d_{b{\bf i}}^{\dag}d_{b{\bf i}}$, 
where $d_{a{\bf i}}$ ($d_{b{\bf i}}$) is an annihilation operator 
for an $e_g$ electron in the $d_{x^2\!-\!y^2}$ ($d_{3z^2\!-\!r^2}$) 
orbital at site ${\bf i}$, ${\bf a}$ is the vector connecting 
nearest-neighbor sites, and $t^{\bf a}_{\gamma \gamma'}$ is the 
hopping amplitude \cite{comment1} given by 
$t^{\bf x}_{aa}$=$-\sqrt{3}t^{\bf x}_{ab}$=$-\sqrt{3}t^{\bf x}_{ba}$=
$3t^{\bf x}_{bb}$=$3t/4$ for ${\bf a}$=${\bf x}$ and 
$t^{\bf y}_{aa}$=$\sqrt{3}t^{\bf y}_{ab}$=$\sqrt{3}t^{\bf y}_{ba}$=
$3t^{\bf y}_{bb}$=$3t/4$ for ${\bf a}$=${\bf y}$ \cite{unit}.
The second term with the energy $J'$ represents the AFM coupling 
between nearest-neighbor classical $t_{2g}$ spins 
normalized to $|{\bf S}_{\bf i}|$=$1$.
The third term describes the coupling of an $e_g$ electron with 
the $(x^2$$-$$y^2)$- and $(3z^2$$-$$r^2)$-type JT modes (dimensionless),
given by $q_{2{\bf i}}$ and $q_{3{\bf i}}$, respectively.
This term is characterized by the static JT energy $E_{\rm JT}$
\cite{adiabatic}.

Intuitively, it can be understood that the competition between kinetic 
and magnetic energies can produce an S-AFM state:
The system with $J'$=$0$ is a two-dimensional (2D) ferromagnetic (FM) 
metal to optimize the kinetic energy of $e_g$ electrons, 
while it becomes a 2D AFM insulator at $J'\!\agt \!t$ to exploit the 
magnetic energy of the $t_{2g}$ spins.
For smaller but nonzero values of $J'$, there occurs a mixture of 
FM and AFM states one example of which is the CE-type AFM structure 
at $x$=$1/2$, schematically shown in Fig.~1(a).
In this S-AFM state, a 1D conducting path can be defined by connecting 
nearest-neighbor sites with parallel $t_{2g}$ spins.
Note, however, that the shape of the optimal path is not obvious.
A path with a large stabilization energy is needed to 
construct a stable 2D structure.
Thus, our purpose here is to specify the shapes of (quasi-)stable 1D paths
in the S-AFM manifold.

Let us start with the case of $E_{\rm JT}$=$0$ 
and no electron correlation, 
which allows us to illustrate clearly the importance of topology 
in the present problem.
In Fig.~1(b), the energies per site analytically obtained 
for various states are plotted as a function of $J'$: 
In a 2D FM metal, the kinetic-energy gain is reduced by $2J'$, due to
the loss of magnetic energy per site. 
In a 2D AFM insulator, the magnetic-energy gain per site is $2J'$.
In S-AFM states, the optimized periodicity $M$ for a 1D path along 
the $a$-axis direction is numerically found to be given by 
$M$=$2/n$, in agreement with Ref.~\onlinecite{Koizumi}, where 
$n$(=1$-x$) is the $e_g$-electron number per site.
Thus, $M$ is set as $4$ at $x$=$1/2$, leading to 
$2^4$ paths, categorized into the four types shown in Fig.~1(b).
The energy corresponding to each path is calculated and it is found that 
the S-AFM state with zigzag3 path is stabilized 
for $0.1t$$\alt$$J'$$\alt$$0.35t$.
In the rest of the paper, this energy diagrams will not be shown,
but a ``window'' for a stabilized S-AFM state 
always exists for $J'$ of the order of $0.1t$ at other densities and 
for $E_{\rm JT}$$\ne$$0$.
As shown in Fig.~1(c), the straight and zigzag1 paths lead to a metal,
while the zigzag2 and zigzag3 paths induce a {\it band-insulator}. 
Since it has a larger bandgap (equal to $t$), the zigzag3 path is
stabilized at $x$=$1/2$.

The S-AFM structure with zigzag3 path is nothing but the well-known 
CE-type AFM state, and 
our analysis predicts that this state is very stable at $x$=1/2.
However, the same analysis for $x$$\ge$$2/3$ does not lead to a zigzag 
path as the optimized one but a straight one, which disagrees with
experiment.
Thus, it is necessary to find a quantity other than the energy to discuss 
the possible preferred paths that may arise from a full calculation
including nonzero $E_{\rm JT}$ and Coulomb interactions.
Reconsidering the results at $x$=$1/2$ lead us to the idea that 
the number of vertices along the path, $N_{\rm v}$, may provide the key
difference among paths.
A confirmation of this idea is given by the calculation of energies for 
the $2^6$ and $2^8$ paths at $x$=$2/3$ ($M$=6) and $3/4$ ($M$=8), 
respectively.
As shown in Fig.~1(d), the energies can be grouped in 
$(M/2$+1)-fold bands, each of which is characterized by 
$N_{\rm v}$(=$0$, $2$, $\cdots$, $M$).
This suggests that $N_{\rm v}$, a topological feature, 
is relevant for the physics of the S-AFM states.

\begin{figure}[h]
\vskip3.1truein
\hskip-2.2truein
\centerline{\epsfxsize=0.9truein \epsfbox{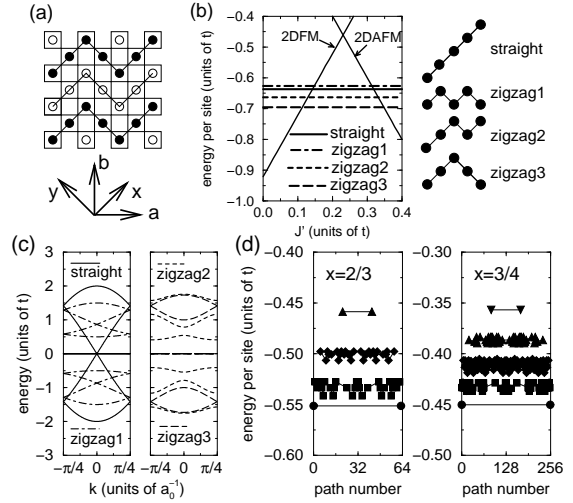} }
\vskip-2.1truein
\caption{(a) Schematic view of the S-AFM structure in the $a$-$b$ plane 
with oxygens at the corners of squares. 
Solid and open circles represent, respectively, up and down $t_{2g}$ spins. 
Solid lines indicate the hopping paths for $e_g$ electrons. 
(b) Energy per site vs. $J'$ for several magnetic arrangements
at $x$=$1/2$ and $E_{\rm JT}$=$0$.
Four types of paths in the S-AFM state are shown.
(c) Dispersion curves for the S-AFM states with straight, zigzag1, zigzag2, 
and zigzag3 path. Here $a_0$ is the lattice constant in the $a$-$b$ plane.
(d) Energies per site for all possible paths at $x$=$2/3$ and $3/4$ 
without JT distortions.
Circles, squares, diamonds, up-triangles, and down-triangles denote 
the results for $N_{\rm v}$=0, 2, 4, 6, and 8, respectively.
}
\end{figure}

Let us include the JT distortion to substantiate our ideas.
By writing the JT modes in polar coordinates as
$q_{2{\bf i}}$=$q_{\bf i}$$\sin \theta_{\bf i}$ and 
$q_{3{\bf i}}$=$q_{\bf i}$$\cos \theta_{\bf i}$,
``phase-dressed operators'', 
${\tilde d}_{a{\bf i}}$ and ${\tilde d}_{b{\bf i}}$, 
are introduced as 
${\tilde d}_{a{\bf i}}\!=\! e^{i\theta_{\bf i}/2} \!
[d_{a{\bf i}} \cos (\theta_{\bf i}/2) \!+ \! 
d_{b{\bf i}} \sin (\theta_{\bf i}/2)]$ 
and ${\tilde d}_{b{\bf i}}\!=\! e^{i\theta_{\bf i}/2} \!
[-d_{a{\bf i}} \sin (\theta_{\bf i}/2) \!+ \!
d_{b{\bf i}} \cos (\theta_{\bf i}/2)]$ with 
$e^{i \theta_{\bf i}/2}$ representing the molecular Aharonov-Bohm effect. 
The amplitude $q_{\bf i}$ is determined by a mean-field approximation 
\cite{Hotta}, while the phases, $\theta_{\bf i}$'s, 
are interrelated through the Berry-phase connection to provide 
the winding number $w$ along the 1D path 
as $w$=$\oint_C$$d{\bf r}$$\cdot$$\nabla$$\theta/(2\pi)$ \cite{connection},
where $C$ forms a closed loop for the periodic-lattice 
boundary conditions \cite{phase}.

Mathematically $w$, a Chern number, is proven to be an integer \cite{Hotta}.
In this system, it may be decomposed into two terms as 
$w$=$w_{\rm g}$+$w_{\rm t}$.
The former, $w_{\rm g}$, is the geometric term, which becomes $0$ ($1$) 
corresponding to the periodic (anti-periodic) boundary condition 
in the $e_g$-electron wavefunction.
The discussion on the kinetic energy leads us to conclude that 
the state with $w_{\rm g}$=$0$ has lower energy than that with 
$w_{\rm g}$=$1$ for $x \! \ge \! 1/2$ \cite{comment3},
in agreement with the two-site analysis \cite{Takada}.
Thus, $w_{\rm g}$ is taken as zero hereafter.

\begin{figure}[h]
\vskip0.4truein
\centerline{\epsfxsize=2.8truein \epsfbox{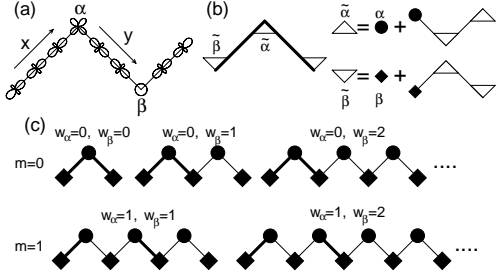} }
\vskip-1.2truein
\caption
{(a) A typical building block for a 1D path for an $e_g$ electron 
with JT distortions. 
(b) General structure of the lowest-energy-state path and
the renormalization scheme for the vertices $\alpha$ and $\beta$.
The thick (thin) line denotes the straight-line part 
with (without) an $e_g$ electron localized on it. 
The solid circle and diamond denote, respectively, the bare vertices, 
$\alpha$ and $\beta$, while open up- and down-triangles indicate 
the renormalized vertices, ${\tilde \alpha}$ and ${\tilde \beta}$. 
Note that the periodicity of the 1D path is given by 
$M$=$2/n$=$2/(1\!-\!x)$. 
(c) Groups of 1D paths derived from mother states with $m$=$0$ and $1$. 
Paths in the first column corresponding to the mother WC structures
with $w$=$2m$+$1$, which produce daughter states with 
$w$=$2m$+$2$, $2m$+$3$, $\cdots$.} 
\end{figure}

To show that only $N_{\rm v}$ determines the topological term $w_{\rm t}$, 
let us consider the transfer of a single $e_g$ electron along the path shown
in Fig.~2(a). 
On the straight-line part in the $x$-($y$-)direction, 
the phase is fixed at $\theta_{\bf x}$=$2\pi/3$ 
($\theta_{\bf y}$=$4\pi/3$), because the $e_g$-electron orbital 
is polarized along the transfer direction.
This effect may be called an ``orbital double-exchange (DE)''
in the sense that the orbitals align along the axis direction 
to make the transfer of the electron smooth, similarly
as the FM alignment of $t_{2g}$ spins in the usual DE mechanism. 
Thus, $w_{\rm t}$ does not change on the straight-line part of the path. 
However, when the electron passes the vertex $\alpha$ ($\beta$),
the phase changes from $\theta_{\bf x}$ to $\theta_{\bf y}$ 
($\theta_{\bf y}$ to $\theta_{\bf x}$), 
indicating that the electron picks up a phase change of 
$2\pi/3$ ($4\pi/3$). 
Since these two vertices appear in pairs, $w_{\rm t}$(=$w$) is evaluated as 
$w_{\rm t}$=$(N_{\rm v}/2)(2\pi/3 \!+\! 4\pi/3)/(2\pi)$=$N_{\rm v}/2$.
The phases at the vertices are assigned as an average of the phases 
sandwiching those vertices, $\theta_{\alpha}$=$\pi$ and
$\theta_{\beta}$=$0$, 
to keep $w_{\rm g}$ invariant.
Then, the phases are determined at all the sites once 
$\theta_{\bf x}$, $\theta_{\bf y}$, $\theta_{\alpha}$, 
and $\theta_{\beta}$ are known.

Now we include the cooperative JT effect, important 
ingredient to determine COO patterns in the actual manganites. 
Although its microscopic treatment is involved, 
we can treat it phenomenologically as a constraint for macroscopic 
distortions \cite{Hotta}, 
energetically penalizing $w$=0 and $M/2$ paths.
In fact, it is numerically found that $w$=1, 2, $\cdots$, $M/2$$-$1 
paths constitute the lowest-energy band and 
they can be regarded as degenerate, since its bandwidth 
is about $0.01t$, much smaller than the interband energy difference 
($\approx \! 0.1t$).
Summarizing, the cooperative JT effect gives us 
two rules for the localization of $e_g$ electrons; 
(i) they never localize at vertices; 
(ii) if an electron localizes at a certain site on one of
the straight segments in the $x$-direction, 
the other localizes on one of the straight segments
in the $y$-direction.

Applying these rules, we obtain a general structure 
for the lowest-energy path as shown in Fig.~2(b). 
Important features are the ``renormalized vertices'', ${\tilde \alpha}$ and 
${\tilde \beta}$, abbreviated notations to represent the set of 
straight-line parts that do not have $e_g$ electrons. 
The winding number assigned to ${\tilde \alpha}$ (${\tilde \beta}$) is 
$1/3$+$w_{\alpha}$ ($2/3$+$w_{\beta}$), where the number of vertices 
included in ${\tilde \alpha}$ (${\tilde \beta}$) is $1$+$2w_{\alpha}$ 
($1$+$2w_{\beta}$). 
Thus, the lowest-energy path is labeled by the non-negative integers
$w_{\alpha}$ and $w_{\beta}$, leading to a total winding number 
$w$=$1$+$w_{\alpha}$+$w_{\beta}$. 
Although the topological argument does not determine the precise
position at which an $e_g$ electron localizes in space, 
it is enough to regard a charged straight-line part as a ``quasi-charge''. 
Since the quasi-charges align at equal distance in the WC structure,
the corresponding path is labeled by $w_{\alpha}$=$ w_{\beta}$=$m$,
with $m$ a non-negative integer. 
By increasing $w_{\beta}$ keeping $w_{\alpha}$ fixed, 
we can produce any non-WC-structure paths with $w_{\alpha}$=$m$ 
and $w_{\beta}$=$m$+$1$, $m$+$2$, $\cdots$ (see Fig.~2(c)).
In this way, the WC structure with $w$=$2m$+$1$ can be considered 
the ``mother state'' 
for all non-WC-structure paths with $w$=$2m$+$2$, $2m$+$3$, $\cdots$, 
referred to as the ``daughter states''.
The states belonging to different $m$'s are labeled
by the same $w$, but a large energy barrier exists for the conversion
among them, since an $e_g$ electron must be moved through a vertex 
in such a process.
Thus, the state characterized by $w$ in the group with $m$, once formed,
it cannot decay, even if it is not the lowest-energy state.

Note that the topological argument works irrespective of the detail 
of $H$, since $w$ is a conserved quantity. 
However, it cannot single out the true ground state, since the 
quantitative discussion on the ground-state energy depends on the 
choice of $H$ and approximations employed. 
In fact, either the BS or WC structure can be the ground state, but 
in view of the small energy difference, their relative energy 
will likely change whenever a new ingredient is added to $H$.
Especially, the Coulomb interactions will be important to decide the 
winner in the competition between these structures \cite{coulomb1}.

Now we analyze the charge and orbital arrangement 
in La$_{1-x}$Ca$_{x}$MnO$_{3}$, in which the experimental appearance 
of the BS structure provides key information to specify the 1D path. 
Since the quasi-charges exist in a contiguous way in the BS structure, 
its path is produced from the mother state with $m$=$0$ (see Fig.~2(c)). 
In particular, the COO pattern in the shortest 1D path is uniquely 
determined as shown in Figs.~3(a)-(c). 
(To depict these figures, we performed a mean-field calculation 
for $E_{\rm JT}$=$2t$, but the essential physics does not depend on either 
the approximation or the choice of $E_{\rm JT}$.) 
At $x$=$1/2$, the path is characterized by $w$=$1$ which is 
the basic mother state with $m$=$0$. 
The COO pattern shown in Fig.~3(a) leads to the CE-type AFM state 
\cite{Murakami}.
Those in the paths with $w$=$2$ and $3$ are nothing but 
the BS structures experimentally observed at $x$=$2/3$ and $3/4$
\cite{Mori}. 
Note that the short-period zigzag part explains the peculiar feature 
exhibiting small oscillations in $q_{\bf i}$ at less- (non-) 
distorted Mn$^{4+}$ sites, as suggested in experiments \cite{Mori}.

It may be assumed that the long-range Coulomb interaction $V$ 
destabilizes the BS structure and transforms it to the WC
structure \cite{coulomb2}, but this is not the case;
for the BS $\rightarrow$ WC conversion with 
the help of $V$, an $e_g$ electron must be on the vertex 
in the path with $w$=$2$ or $3$ (see Figs.~3(b) and (c)). 
This is against rule (i) and thus, the BS structure, once formed, 
is stable due to the topological condition, 
even including a weak repulsion $V$.

In the group of $m$=$0$, the WC structure appears only in the path 
with $w$=$1$.
Thus, the WC-structure paths with $w$=$1$ at $x$=$2/3$ and $3/4$ 
are obtained by simple extension of the straight-line part 
in the path at $x$=$1/2$ (see Figs.~3(d) and (e)).
The detailed charge distribution inside the quasi-charge segment is 
determined by a self-consistent calculation with the JT effect, 
leading to the WC structure.
(A similar result can also be obtained for a weak $V$.)
Even if the non-WC structure occurs for $w$=$1$, it is unstable 
in the sense that it is easily converted to the WC structure,
because no energy barrier exists for an $e_g$ electron shift along 
the straight-line part.

\begin{figure}[h]
\vskip0.6truein
\hskip-0.2truein
\centerline{\epsfxsize=3.4truein \epsfbox{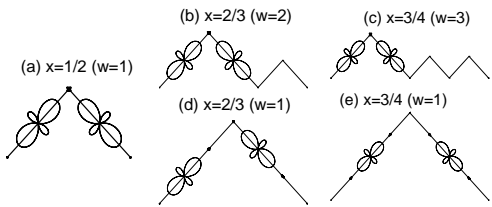} }
\vskip-2.3truein
\caption
{(a) Path with $w$=$1$ at $x$=$1/2$ for $E_{\rm JT}$=$2t$. 
At each site, the orbital shape is shown with its size 
in proportion to the orbital density. 
(b) The BS-structure path with $w$=$2$ at $x$=$2/3$. 
(c) The BS-structure path with $w$=$3$ at $x$=$3/4$. 
(d) The WC-structure path with $w$=$1$ at $x$=$2/3$. 
(e) The WC-structure path with $w$=$1$ at $x$=$3/4$.} 
\end{figure}

The above topological analysis show that 
(1) the WC structure is made of $w_{\rm WC}$=$1$ zigzag paths
and that (2) the BS one contains 
a shorter-period zigzag path with $w_{\rm BS}$=$M/2\!-\!1$=$x/(1\!-\!x)$.
Note that on the BS path, the less-distorted Mn$^{4+}$ sites occupy all 
the vertices ($N_{\rm v}$ equals the number of Mn$^{4+}$ ions), 
while the heavily distorted Mn$^{3+}$ sites appear in pairs 
(the number of Mn$^{3+}$ ions equal to $2$). 
Thus, $w_{\rm BS}$ is rewritten as 
\begin{equation}
\label{w2}
w_{\rm BS} = {N_{\rm v} \over 2} 
={{\rm Number~of~Mn}^{4+}~{\rm ions} \over 
{\rm Number~of~Mn}^{3+}~{\rm ions}} ={x \over 1-x}.
\end{equation} 
Since $w_{\rm BS}$ is an integer, we can predict that at specific values 
of $x$[=$w_{\rm BS}$/($1$+$w_{\rm BS}$)], such as $1/2$, $2/3$, $3/4$, etc.,
nontrivial charge and orbital arrangement will be stabilized in agreement 
with the experimental observation \cite{Mori}.

Two comments are in order. 
(1) To understand the observed BS structure, we focussed on 
paths produced from the $m$=0 mother state, but 
we could have used a $m \ne 0$ mother state.
Thus, we anticipate the existence of a relation similar to Eq.~(\ref{w2}) 
in a higher hierarchy (with much longer periodicity), 
leading to a devil's staircase structure in the 1D path. 
(2) The S-AFM structure with the zigzag path for $w$=$1$ 
explains the AFM phase observed in experiment at $x$=$2/3$, 
but we further predict a similar S-AFM state at $x$=$3/4$ in the WC
structure. 
Our theory also predicts a peculiar spin pattern in the BS structure. 
Observation of such structures will provide a stringent test for the 
validity of our topological scenario.

In summary, the BS and WC structures in the manganites have been 
classified using the winding number $w$ associated 
with the Berry-phase connection along the zigzag 1D path. 
Predictions are made for novel spin and orbital states in heavily doped
manganites.

T.H. thanks S. Yunoki and C. H. Chen for discussions. 
T.H. and Y.T. are supported from the Ministry of Education, Science, 
Sports, and Culture of Japan. 
E.D. is supported by grant NSF-DMR-9814350.


\end{multicols}

\begin{references}
\bibitem{unit} Here $t$ is defined as the hopping amplitude between 
nearest-neighbor $d_{3z^2\!-\!r^2}$ orbitals along $z$-axis.
\bibitem{adiabatic} Note that the JT distortions are treated adiabatically.
\bibitem{phase} Note that the concept of winding number $w$, 
which implicitly assumes mobile electrons, is relevant even for 
an insulator, provided that it is a band insulator, because 
the system is described by the superposition of extended 
wavefunctions with which we can associate $w$.
\bibitem{coulomb1}
Qualitatively, the present result will not be altered by the 
on-site Coulomb repulsion, since the double occupancy of electrons at
a site is already suppressed by the JT stabilization mechanism.
\end{references}
\end{document}